\documentclass[preprint,tightenlines,eqsecnum,aps,epsfig]{revtex4}

\usepackage{graphicx}
\usepackage{epstopdf}
\usepackage{epsfig}
\newcommand \Pomeron {I\!\!P}

\def\beq{\begin{equation}}   \def\eeq{
\end{equation}}

\begin{document}
\title{The energy dependence of the
 hard exclusive
diffractive processes in pQCD as the function of
momentum transfer.}

\author{ B. Blok\email{Email :blok@physics.technion.ac.il} }
\affiliation{Department of Physics, Technion---Israel Institute of Technology, 32000 Haifa, Israel}
\author{ L. Frankfurt\email{E-mail: frankfur@tauphy.tau.ac.il} }
\affiliation{School of Physics and Astronomy, Raymond and Beverly
Sackler Faculty of Exact Sciences, Tel Aviv University, 69978 Tel
Aviv, Israel}
\author{ M.Strikman\email{\E-mail: strikman@phys.psu.edu}}
\affiliation{Physics Department, Penn State University, University Park, PA, USA}
\thispagestyle{empty}
\begin{abstract}
We predict   the dependence on energy of photo(electro) production processes:
$\gamma(\gamma^*)+p\rightarrow V+ X$ with  large rapidity gap at small x  and large
momentum $-t$ transferred to $V$ in pQCD.  Here V is a heavy quarkonium  ($J/\psi, \Upsilon$) or
longitudinally polarized  light vector meson (in the electroproduction processes), etc. In the
kinematics of HERA  we calculate the dependence on energy of cross sections of these
 processes as the function of momentum transfer $t$, photon virtuality $Q^2$ and/or quarkonium mass.  In
the kinematical region  $Q_0^2\le -t\ll Q^2+M^2_V$   the
nontrivial energy  dependence of the
cross section for the vector meson production due to the  photon
scattering off a parton follows within QCD from the summing of the
double logarithmic terms.      In the second regime $-t\ge
Q^2+M^2_V$  within DGLAP  approximation in all orders of perturbation theory the
$q\bar q - {\rm parton}$ elastic  cross section is energy independent. We show
that the correct account of the double logarithmic terms and of
the gluon radiation including kinematical constraints removes the
disagreement between pQCD calculations and recent HERA experimental
data. The explicit formula for
the
dependence of
the differential cross section$\displaystyle{\frac{d^2\sigma}{dtdx_J}}$ of these processes on $s_{\gamma^*N}$ is
obtained.
  We show that perturbative  Pomeron type behavior
may reveal itself only at energies significantly larger than those available at HERA.
 In addition we evaluate the energy dependence of DVCS processes.
\end{abstract}

\maketitle \setcounter{page}{1}
\section{Introduction.}

\par
The hard inelastic exclusive photo(electro) production processes:
$\gamma^* p\rightarrow V +{\rm rapidity ~ gap} +X$  with a large  rapidity gap and
proton dissociation  into hadrons draw  a lot of attention recently  (V is vector $\rho-$ meson,
charmonium $J/\psi$, or bottomium). (Such processes can be  dubbed hard inelastic
diffractive (HID) processes, in order to distinguish them from the diffractive
processes, when the final particle remain a proton: $X=p$). These
processes  with $V=J/\psi$ were  extensively studied experimentally at HERA
\cite{ZEUS,H1,galina} due to a reasonably large cross section and a clean
final state - production of an isolated charmonium state with large
transverse momentum which is separated from the proton remnants by
a large rapidity gap. Theoretically this process is advantageous,
since the large charmonium mass  $m^2_V$ makes it possible to
evaluate cross section of  the process in perturbative QCD,  for
different values of  transverse momentum transfer $-t$ and
virtuality of the photon $Q^2$ \cite{AFS, Forshaw1, Levin}.
Moreover, the QCD factorization theorem has been proved \cite{BFGMS,CFS} for
diffractive photo (electro) production  of vector mesons at least for the case of the
longitudinal photon
putting the calculations of diffractive cross sections  on the solid
theoretical basis. Processes where proton remains intact  are strongly suppressed at
large $t$ by the nucleon form factor and will be considered briefly only.

\par
Moreover,  the charmonium photoproduction is to be probed soon at ultraperipheral
collisions at LHC \cite{Mark} for significantly larger energies. The CMS and ATLAS
detectors are well suited for such an investigation since  they cover large rapidity intervals. The
ALICE detector may be  capable of studying this process in a certain rapidity range as  well.

\par
Phenomenological  analysis of the H1 and ZEUS  data on diffractive charmonium
production \cite{ZEUS,H1}, carried out in  ref. \cite{FSZ} within framework
of the QCD factorization theorem has demonstrated that the experimental data can be described
if  the energy dependence of  two body pQCD scattering amplitude would be
significantly slower at large $-t$, than at $t=0$. Recently, a new
experimental data on the  photoproduction process
$\gamma p\rightarrow J/\psi+{\rm rapidity ~gap} +X$ was
reported \cite{galina}. This data also indicate a rather dramatic
slowing down  of the dependence of the cross section of this
process on the energy with increase of $-t$. This  conclusion
is consistent within the experimental uncertainties with the  phenomenological
 analysis of ref. \cite{FSZ}. The observed behavior  differs \cite{galina}  from the
 predictions
which assumed that the  energy dependence of the
amplitudes describing two body processes $f(s, M^2_{X}, t, Q^2)$ can
be evaluated in terms
of the BFKL approximation with the coupling constant independent of  t \cite{Forshaw1,Levin},
or with $\alpha_{eff}=\alpha_s(t)$
\cite{Lipatov}.

\par
The novel QCD effect discovered in this paper is that the pattern of energy dependence of
cross section of the HID  on $t$, observed at HERA, follows    from the DGLAP  approximation
of pQCD. The  amplitude of the hard two body collisions has a rather simple  form in a wide
interval of high energies where the double logarithmic (DL) approximation is legitimate:
\beq
f = f_{DL}(x_1,x_2, M^2_X, t, Q^2, M^2_V)F(x, M^2_X, t).
\label{f}
\eeq
Within the DGLAP approximation \cite{DGLAP1,DGLAP2,DGLAP3}  the function $f_{DL}$  is obtained
by summing large  $\alpha_s(N_c/2\pi) \ln(x_J/x) \ln(Q^2+M^2_V)/(Q^2_0-t)$ terms,
that arise in the integration over parton transverse momenta in the domain where
$\alpha_s\ll 1$.  Here $x_J=-t/M^2_X$ is the fraction of the proton momentum carried by the parton
involved in the large $t$ elastic scattering.
Variables $x_i$ are the fractions of the proton momentum carried by gluons exchanged
in $t$ channel which are attached to a parton  of the proton
(For the explicit definition of $x_1, x_2$ see section 2.).  Since at large $t$
transverse momenta of exchanged gluons are large, $x_i$ are not vastly different.

\par
 Note  that it has been demonstrated  in ref. \cite{Forte}, see also refs. \cite{DGLAP2,DDT},
that  increasing with energy DL terms provide a good description of  the structure functions
of proton measured at small $x$ at HERA. This observation suggests that at
 HERA energies $F=1$. (  We normalise $f_{DL}$ and $F$ in such a
 way that $F={\rm constant}=1$ at low energies.)

\par
The function $F=1$ within the LO DGLAP approximation to pQCD, but it is
a function of energy within the BFKL and resummation models
\cite{BFKL,Ciafaloni,ABF}. This is because these approaches take
into account the contributions that increase with the energy but
do not contain double logarithmic (DL) terms. We estimate energy
dependence of $F$ in the paper. The effect is small at HERA
energies but may become noticeable  at significantly higher
energies.

 \par
 In the kinematic region  $Q_0^2<-t< Q^2+M^2_V$ the energy dependence of
hard amplitudes is  determined by $f_{DL}$. The equation \ref{f}  differs in this
kinematic range from the Pomeron
exchange expression $(s/s_0)^{\alpha_{\Pomeron}(t)-1}$  often
used to describe the data. In particular,  DL approximation predicts strong dependence of phenomenological
$\alpha_{\Pomeron}(t)$ on $Q^2$ and on incident energy (i.e. non universality, dependence on the external
conditions). This non universality of $\alpha_{\Pomeron}$
specific for DGLAP approximation has been observed at HERA at $t=0$ \cite{HERA}:
\beq
F_2(x,Q^2)\sim \exp(r(Q^2)\log(1/x)), r(Q^2)\sim 0.05\log(Q^2/\Lambda^2).
\label{X}
\eeq

\par
In the kinematical region $-t\ge Q^2+M^2_V$,  $f_{DL}$ is  energy independent.
Hence the energy dependence of the cross section is determined by the function $F$
which does not contain large logarithms from the integration over parton transverse
momenta and 
which
 is equal to one at  HERA energies.  Thus pQCD predicts a sharp decrease
with $-t$
of the rate of the rise with energy of the photo(electro) production cross sections
 as compared to forward scattering at HERA energies.

 The estimates of the kinematic range  dominated by double logarithms
 \cite{Ciafaloni,transition} show that the universal Pomeron behavior
may be valid only  for the energies well beyond
 the kinematical region: $\ln(x_J/x)\sim (2\div 3)  \log(Q^2/(-t+Q^2_0))$ occupied by double logarithms, i.e.
when the DL terms disappear. At $Q^2+M^2_V\ge 10 GeV^2$  this condition corresponds to the kinematics
far beyond the kinematical  range of HERA.
The explicit analysis of the phase-space constraints on the multi Regge kinematics
due to the  energy-momentum
conservation shows that even at ultraperipheral processes at the LHC (where
$x $ down to $10^{-6}$ can be reached) these constraints limit the possibility of the onset of the
Pomeron behavior to the kinematic range $-t\ge Q^2+M^2_V$.
It will be very interesting to look for  onset  of a Pomeron
 behavior in pQCD in ultraperipheral collisions in this particular
kinematic region.

\par
Let us note that the dominance of the DL terms at small $x$ in a
wide kinematical range  is a common feature of  DGLAP, one of
saddle points in the improved BFKL, and the  resummation  models
(see the discussion  in refs. \cite{Ciafaloni,ciafaloni1,ABF}). In
the improved BFKL approach \cite{lipatovfadin} double logarithmic
terms are accounted properly in the same way as  in the
resummation models. Thus the energy dependence given by$f_{DL}$
should disappear with increase of $-t$ in all approaches as long
as the contribution of the BFKL saddle point can be neglected. In
other words, our results will remain quantitatively correct also
in these approaches if one treats correctly the collinear terms
and includes the energy momentum conservation constraints. The
difference may reveal itself only at the energies far beyond HERA.
On the contrary the contribution of the BFKL saddle point
\cite{BFKL} in the inverse Mellin transformation has no DL terms
and therefore it does not lead to a  peculiar dependence on $t$ of
the rate of increase of the cross section with  energy discussed
in the paper.

\par
We solve the DGLAP evolution equations in the form of the DL  approximation
in a large interval of $t$. It is known that the DL approximation  gives quantitative
description of  the HERA data on the structure functions of a proton
\cite{Forte}. Technically, the main effect related
to the  increase of  $-t$ is  the following. In the case of $t=0$
the strong energy dependence of QCD cross section arises because
the energy dependent terms $\log (x_0/x)$ are enhanced by  the
factor $\log(\log(Q^2/\Lambda^2)/\log(Q^2_0/\Lambda^2))$ where
$Q_0^2$ is an initial virtuality.  This factor originates from the integration over transverse momenta in the range
$Q_0^2\ll k^2_t\ll Q^2$. For $-t\gg Q^2_0$  the effective range of
integration changes to $-t\ll k^2_t \ll Q^2+M^2_V$, leading to the
decrease of the rate of the energy increase with the increase of
$-t$. At $-t\sim Q^2+M^2_V$ the DL terms  disappear all together.
Then the energy dependence may arise due to either nonperturbative
initial condition for QCD evolution  or the terms that do not contain integration over large transverse momenta.

\par Since the intermediate state invariant mass of the $c \bar c $ system at large $-t$  is significantly larger then the masses of charmonium
states we can  approximate $f$ at $t\ne 0$ with unequal $x_i$   by the nonforward parton
distribution with $x=(x_1+x_2)/2$. It was shown that at small $x$ and large scale controlling hardness of the process
this approximation does not introduce significant uncertainties \cite{BFGMS,FS2,Freund}.

\par
The nonforward parton distributions for diffractive Z boson
production were studied also in ref. \cite{Bartels}  in the
kinematical  region:  $\log(Q^2/Q_0^2)>\log(x_0/x)$ which differs
from the kinematical region of HID
($\log(Q^2/Q_0^2)\ll \log(x_0/x)$) considered in this paper. Cross
section of this process is strongly suppressed at large t as
compared to that for HID by the square of nucleon form factor.
Moreover  impulse approximation=QCD factorization should be
modified at large t where two gluon scattering off two different
partons becomes preferable. Besides, the dependence on the running
coupling constant derived in our paper within DGLAP approximation,
differs from the one suggested in ref. \cite{Bartels}.

\par
The onset of the black disc limit for the processes with large rapidity
gap  may occur in the kinematics where $x_J=-t/M^2_X$ would be
sufficiently small, i.e. in the kinematics where the gap between $V$
and the system $X$ is relatively small.
The distinctive signature of
this regime is a significantly slower decrease of cross section with $-t$ as compared to pQCD regime.
Note also that in the black regime inelastic diffraction is suppressed. In
the discussed process this will lead to a further slowing down of the
increase of the cross section with energy at fixed $x_J, t$ or perhaps even  to a decrease of the cross section with energy.
This effect is amplified with increase of $x_J$ since the transverse distribution of gluons becomes more localized in the transverse plane.
\par
The paper is organized in the following way. In  Sec. II we discuss the
kinematics of the HID  process.  In Sec. III we consider the DGLAP evolution
equations for the nonforward GPD  and solve them in the double logarithmic
approximation.   In Sec. IV we evaluate the energy dependence of the amplitudes of
hard diffractive processes  for the kinematics achieved at  HERA .
 In Sec. V  we discuss briefly dependence on energy of the cross section of DVCS
process as a function of $t$. In the conclusion we discuss the directions for the
future progress.

\section{Kinematics of the diffractive production at high energies.}

\par The theoretical framework for the calculation of inelastic
diffractive processes in the high energy processes has been developed in the
seventies and it is the so called  triple-Pomeron limit \cite{Gribov}.  The original
approach used independence of Pomeron trajectory on external conditions - the property
of the soft QCD.  It has been understood later \cite{FS3}  how these ideas
can be adjusted to evaluate the cross sections of the hard processes with large rapidity
gap directly in pQCD where the QCD factorization substitutes the Pomeron pole
factorization. The analogue of the triple "Pomeron" vertex is calculable within the
pQCD (see Fig.~1 and discussion below). The natural variables in these
processes are the the invariant mass of hadronic states produced in the proton
dissociation  $M_X$, the square of the transverse momentum transfer
$-t=-(p_\gamma -p_V)^2$ and the square of the invariant energy for
$q\bar q ~{\rm parton}~ j$ elastic scattering:
\beq
s'\equiv s_{q\bar q+ "{\rm parton}~ j"}=x_Js - Q^2, \label{k1}
\eeq
$s=W^2_{\gamma p} $ is an invariant energy squared of a full proton-photon system.
These quantities are connected by the kinematic relation
 \beq
x_J=-t/(M^2_X-m^2_p-t),
\label{int}
\eeq
  The rapidity gap for sufficiently large momentum transfer -t and invariant energy s  is given by
\beq
 \delta y=\ln \frac{s}{\sqrt {(M_V^2-t)(M_X^2-t)}}.
  \label{k3}
 \eeq
 Finally, there exists a kinematical boundary on $-t$. For the forward scattering
 in the kinematics where $M^2_X\gg Q^2$ :
  \beq
  -t_{\rm min}=(M^2_X-m^2_p)(M^2_V+Q^2)/2s.
   \label{k51}
   \eeq
This boundary is however irrelevant from the practical point of
view, since in the kinematics characteristic for the processes with large
gap in rapidity $M_X^2/s \ll 1$.  Therefore in the essential kinematical domain
investigated in this paper:   $|t|\gg |t|_{min}$ and corrections due to $t_{min}$ can
be neglected.

Note that though in principle
$M_X$ can be measured  using energy-momentum conservation in terms
of the momentum carried by vector meson, in practice
it can be determined
from the measurement of  the rapidity interval occupied by system $X$.

Cross section of the HID processes $\displaystyle{\frac{d\sigma}{dtdx_J}}$ is
calculable at large $-t$ within pQCD as the consequence of the QCD factorization
theorem cf. \cite{AFS,FSZ,Forshaw1} :
 \beq
\frac{d\sigma}{dtdx_j}=\frac{d\sigma_{\gamma+ j\to
V+j}}{dt}\left((81/16) G(x_J,t)+\sum_i (q_i(x_J,t)+\bar q_i(x_J,t))\right).
\label{k5}
\eeq
The  cross section is the product of two factors, both of
which
dependent on $W_{\gamma p}$: $G(x_J,t)$ and the cross section of photon scattering on a parton $j$ given
by a nonforward amplitude $f$  (see Fig.~1) (Here $\displaystyle{\frac{d\sigma_{\gamma+ j\to
V+j}}{dt}}$ is the cross section of scattering off a parton $j$ ). In order to evaluate
the energy dependence of the cross section one  needs to take into
account the energy dependence  of both factors in eq.~\ref{k5}.

\par
The first
factor in eq.~\ref{k5} is
\beq
\frac{d\sigma}{dt}=\frac{\vert f\vert ^2}{16\pi }.
\label{p2}
\eeq
Here the amplitude $s'f$ is a hard amplitude of a photo/electro production of  a
system $V$ when a virtual photon scatters off a parton with 4-momentum $x_Jp$.
The amplitude f is a convolution of impact factor describing transition $\gamma^*\to V$
with the generalized  parton distribution $D$ where the initial condition for the  QCD
evolution is the amplitude of the  scattering off a single parton. This amplitude
depends on four parameters $-t$, $Q^2, M^2_V$ and the effective energy of
parton-photon system (eq.~\ref{k1}).
In addition  $D$ depends on two arguments $x_i$  rather than on $x_J$:
\beq
x_1=(M^2+Q^2)/(s'+Q^2);
x_2=(M^2-M^2_{V})/(s'+Q^2).
\label{arg}
\eeq
Here $M^2$ is  the invariant mass of intermediate state in the impact factor. In the
charmonium production $M^2$ is approximately the invariant mass of $c\bar c$ pair:
\beq
M^2=\frac{k^2_t+m^2_c}{z(1-z)},
\label{pe}
\eeq
and $z$ is the fraction of the
 photon momentum carried by one of the charmed quarks,  $m_c$
is the running charmed quark mass. $k_t$ is typical momentum of the charmed quark.
To account for nonzero $t=-\Delta^2$  one should substitute in the above formulae
$k_t$ by $k_t-z\Delta$.
As the consequence of large mass of $c$ quark the contribution
of the end point ($z\sim 0, z\sim 1$) is negligible. So essential $z\approx 1/2$.

Since  $M^2$ is significantly larger than $M^2_V$  for  large $\Delta^2$,
  $x_i $ are not very different.  So it can be  demonstrated following ref. \cite{BFGMS,
Freund}  that  to a good approximation $D$ is equal to diagonal
parton distribution, but with  $x=(x_1+x_2)/2$:
 \beq
x_1D(x_1,x_2,t,Q^2,Q^2_0)\approx [(x_1+x_2)/2] D((x_1+x_2)/2, t,
Q^2, Q^2_0). \label{tr1} \eeq \beq x\sim
(2M^2-M^2_V+Q^2)/2x_Js.\label{d1} \eeq Technically this  result is
due to the fact that the initial condition for the generalized
parton distribution  for small x practically coincides with a
diagonal one  with $x=(x_1+x_2)/2$ \cite{Freund}. The main
contribution to skewness comes from the cell in the Feynman
diagram of Fig. 1 closest to the vector meson. It is easy to check
that this is a property of  the DGLAP dynamics in the kinematics
$-t\ll Q^2+M^2_V$.  For $-t\ge Q^2+M^2_V$ there are no  logarithms
from the integration over transverse momenta and nonforward
distributions discussed below do not depend on energy in the DGLAP
approximation. Thus the energy dependence of $f$ and nonforward
diagonal distribution are the same.

\par
The study  of the first factor in eq.~\ref{k5}  will be the central subject of our paper.
The total cross section is given by the integral
\beq
{d\sigma\over dt} =\int_{\rm kinematical~ cuts} dx_j
{d\sigma\over dtdx_j}.
\label{k7}
\eeq

 \par
We carry on our calculations for photoproduction in the kinematic region
$-t_{\rm min}\ll-t<M^2_V$ and for electroproduction in the kinematic range
$-t_{\rm min}\ll-t\ll M^2_V+Q^2$. We shall see below that for complimentary kinematic
range of $-t > M^2_V+Q^2$ the energy dependence is quite different.

\section{DGLAP evolution equations for nonzero $-t$.}
\subsection{General structure.}

\par
Let us now consider the nonforward parton distribution that  enters
in eq.~\ref{k5}. In this section we calculate it in  the LO DGLAP  approximation
and explain how to generalize these results to the   NLO approximation of pQCD.  It
was explained above that the generalized parton
distribution can be approximated at small $x$  by the parton distribution with  the argument
$x=(x_1+x_2)/2$ and  with the  transverse momentum transfer  $t=-\Delta^2$. We now
proceed to calculate this distribution.  First,  it is easy to show that in the class of gauges
$(C A)=0$, where   $C^{\mu}=c_1p^{\mu}+c_2q^{\mu}$ ($c_i$ are  numbers of the order 1), used
in ref. \cite{DGLAP2,DDT}  only ladder diagrams depicted  in Fig.~2   contribute to the parton distribution.
(This is not true in other gauges,  in particular  in the Feynman gauge).   To achieve similarity with the parton model description
of this ladder  we will take in the
paper $c_2=0$. Throughout  this section we will  consider the partons in the ladder to be
massless and $Q^2$ is parameter characterizing hardness of the process:
photon virtuality and/or large mass of quarkonium.

\par
The  calculation of the nonforward DGLAP ladder goes in the same way as for forward
DGLAP ladder. In order to find the contribution of   a given ladder
cell $i$ we use the Sudakov variables: \beq k_i=\alpha_iq+\beta_i
p+k^i_t \label{sud} \eeq The gluon  propagator in the gauge $CA=0$
is given by
\beq D^{ab}_{\mu\nu}(k)=\delta^{ab}\frac{1}{k^2+i0} \left[
g_{\mu\nu}-\frac{C^\mu k^\nu+C^\nu k^\mu}{Ck}+k^\mu k^\nu
\frac{C^2}{(kC)^2}\right]. \label{prop2} \eeq
This gauge is known to be
free of ghosts.  Moreover, it was shown in ref. \cite{DDT} that in
the leading logarithmic approximation (LLA)  the contribution of  the term  $k^\mu k^\nu$
is zero. As a result, since we are working in a LLA we can use the
propagator \ref{prop1}:
\beq
D^{ab}_{\mu\nu}(k)=\delta^{ab}\frac{1}{k^2+i0} \left[
g_{\mu\nu}-\frac{C^\mu k^\nu+C^\nu k^\mu}{Ck}\right]. \label{prop1} \eeq
The proof \cite{DDT} is valid for $-t_{min}\ll -t\ll Q^2+M^2_V$
In the kinematical region $ s\gg -t\gg Q^2+M^2_V$ this prove should
be valid as well since
the contribution of the new structures $\propto \Delta_{\mu}$
is suppressed by the power of energy.

\par
We now perform the standard algebraic calculations using the
propagator \ref{prop1} for the cell $i$.  In the DGLAP
approximation  $k^{2i}_t\gg k^{2(i-1)}_t$, and the integrals over
longitudinal and transverse  momenta decouple. The   use of the
gauge invariance allows to demonstrate that only one tensor
structure, $g_{\mu\nu}^t\equiv g_{\mu\nu}-(p^\nu q^\mu+p^\mu q^\nu)/pq $
leads to the contribution containing terms $\propto \ln(Q^2/Q^2_0)$.
There are additional tensor structures that appear for the nonforward
ladder  but they do not lead to the logarithmic contributions in
the integral  over the
transverse momenta.  Moreover, integrals over transverse momenta
of gluons exchanged between different cells do not produce
$\ln(Q^2/Q^2_0)$. Thus the same structure
leads to  large
logarithms in both forward $(t=0) $ and nonforward cases.

Moreover, the direct calculation shows that the $t$
dependence  is present in the integrand only in
  transverse momenta of propagators in the cell. Hence
  the   DGLAP kernels $\Phi (z)$ do not
depend on $t$. The calculation of integral over $\alpha_i$ is done
by taking residues.  As a result the contribution of a single cell
$i$ in the DGLAP ladder can be written in the form \beq
dP_i=\alpha_s(Q^2)dz\Phi(z) I(\Delta )g_{\mu\nu}^t.\label{o1} \eeq
Here $z_i=\beta_i/\beta_{i-1}$.
In deriving eq.\ref{o1} we took into  account  the
gauge invariance, and
summed the $s$ and $u$ channel
contributions (that gives a factor of 2).  The integral over
transverse momenta $I(\Delta)$ has the form:
\begin{eqnarray}
  I(\Delta)&=&\frac{1}{2\pi} \int d^2k_t \left[\frac{1}{(\vec k_t-\vec \Delta)^2}+\frac{1}{k^2_t}
-\frac{\Delta^2}{k^2_t(\vec\Delta-\vec k_t)^2_t}
\right] \nonumber\\[10pt]
&=& \log (Q^2)/(Q^2_0-t))+{\rm non~ logarithmic~
terms},\nonumber\\[10pt]
\label{tr} \end{eqnarray} where $t=-\Delta^2$. The integral over
$k^2_t$ is
over
 the range $Q^2_0$ to  $Q^2$. However the presence of
large $-t$ leads to a cancellation of the  contribution of small
$k^2_t\le -t$. Hence  the contribution of a single cell  $i$
 to  the ladder has the form:  \beq
dP_i=\alpha_sdz\Phi(z) \log (Q^2/-t+Q_0^2)g_{\mu\nu}^t,
\label{single} \eeq
instead of
\beq
dP_i=\alpha_s dz\Phi(z)\log(Q^2/Q^2_0)g_{\mu\nu}^t,
\label{single1}
\eeq
for the DGLAP ladder with $t=0$. Here $\Phi (z)$ is the conventional
DGLAP kernel. The full answer for the ladder is the convolution:
\beq
\sum_n(\frac{\alpha_s \log(Q^2/(Q^2_0-t))}{4\pi})^n \prod
... \int \frac{d\beta_i}{\beta_i}\Phi(z_i))...
\label{full1}
\eeq
The main difference (besides the definition of $x$  -cf. Eq.
\ref{d1}) between nonforward and forward distributions is the
change of the argument of the logarithm arising from the
integration over transverse momenta.

\par
Note  that the latter  expressions were derived in the  kinematics
$Q^2\gg -t\gg Q_0^2$,
while
in the kinematical region  $Q^2\sim -t$ one
cannot single out the logarithmic contributions since they cannot be
considered any  longer as a large parameter. This means that the upper
limits of integration in the integral of eq.~\ref{tr} must be
taken in such a way that for $-t\sim Q^2$ the argument of the
logarithm is equal to one.  Such choice is allowed since we work
in the LLA and all integrals are calculated with logarithmic
accuracy. In addition, the expression for
the
nonforward
distribution should smoothly match with the forward distribution
as $-t\rightarrow 0$. Thus the arguments of the logarithms  that
appear in the nonforward DGLAP distribution are indeed
$\log(Q^2/(Q_0^2-t))$.
\par
We calculated only the imaginary part of the nonforward ladder. The real part
is also nonzero and can be reconstructed using the dispersion relation over the energy
or more rapidly using  the Gribov-Migdal formula \cite{GM}. However, the contribution
of the real part of $f$ into cross section is numerically small and thus can be neglected.

\subsection{Account of the running coupling constant.}

\par
In order to obtain the  full  answer for the DGLAP approximation
we also need  to account for the radiative corrections to the
propagators and vertices.  An effective method to  account for
these corrections is to  use the dispersion relations over the
mass of the produced parton  in  the same way as in ref.
\cite{DGLAP2,DDT,DMW}. Our interest is in the kinematical domain:
$-t=\Delta^2 \ge Q^2_0$ otherwise dependence on $t$ can be
legitimately neglected. The integrand in eq.~\ref{tr} has two
potentially important kinematical regions:   $k^2_t\le \Delta^2$
and $k^2_t\gg \Delta^2 $. In the first kinematic region the
integrand is strongly suppressed  and it is of order $1/\Delta^2$.
The reason is that the IR singularities in the integrand are
cancelled out when $\vec k_t\rightarrow 0,\vec k_t\rightarrow \vec
\Delta $. As a result, the entire logarithmically enhanced
contribution in the integrand comes from the second kinematic
region. However, in this region \beq
\frac{2(k_t,k_t-\Delta)}{(\vec k_t-\vec\Delta_t)^2(k_t^2)}=
\frac{1}{(\vec k_t-\vec \Delta_t)^2}
+\frac{1}{k_t^2}-\frac{\Delta_t^2}{k_t^2(\vec\Delta_t-\vec k_t)^2}
\sim \frac{2}{ k^2_t} +O\left(\frac{-t}{k^4_t}\right). \label{dub}
\eeq The factor $(k_t, k_t-\Delta)$ follows from gauge invariance
and accounts for the gluon polarizations \cite{Gribov}.  This
means that within the logarithmic accuracy the integral over
transverse momenta is the same as in the forward case, the only
difference is that the integration starts at $k^2_t=-t$. Hence we
can directly use the results of ref.\cite{DDT}. The
renormalization of the ladder is given by diagrams of Fig.~3.
Consider first the contribution of the exchanged gluon and two
effective vertices. Taking the sum of discontinuities over two
vertices and an exchanged gluon we obtain for the integral over
transverse momenta
 \beq (1/\pi)\int (d^2k_t/k^2_t)dk^2_2 Im
\frac{\Gamma(k_2,k)\Gamma(k_2,k-\Delta)d_G(k_2)}{k_2^2+i\epsilon}.
\label{run2}
 \eeq
Here as usual $d_G, d_F$
account for  the multiplicative renormalization of the
gluon propagator
 \beq
 1/k^2\rightarrow d_G(k^2,\sigma)/k^2,
 \label{g1}
 \eeq
 and renormalization of the propagator of the  fermion
 \beq
 1/k^2\rightarrow d_F(k^2,\sigma)/k^2.
\label{g2} \eeq
 (Above we
omitted the additional terms proportional to $\delta(k^2_2)$ since
they are not important for large $t$.)  Therefore we can use the
same  approach as in ref. \cite{DDT}.  Namely, we substitute
$k^2_t\rightarrow -k^2(1-z)$ and use  kinematic identity:
 \beq
k^2=-k^2_t/(1-z)+zk^2_2/(1-z).
\label{id}
\eeq
Then the integral over $k^2_2$ in the dispersion relation \ref{run2}
is determined by the pole at $k^2_2=-k^2_t (1-z)/z$. Thus  in the
Feynman integral  for a renormalized cell we must put
$k^2_2=-k^2_t(1-z)/z$.
\par
Since all renormalized vertices are explicitly written  in ref. \cite{DDT},
the only new element to take into account is that the  left
and the right vertices and  the propagators $d_F$ depend now on different
arguments. However since $k^2_t\gg \Delta^2$ integration does not
introduce any changes relative to the case  $-t=0$.  The factors
$\alpha_s(Q^2), d_G,d_F$ all combine together to  the running
coupling constant $\alpha_s(k^2_t)$.  Then the integral over
transverse momenta  in the ladder is given by
 \beq
\int^{Q^2}_{-t}d^2k_t\alpha_s(k^2_t)/k^2_t\equiv
\chi^{'}=\frac{1}{b}\log(\frac{\log(Q^2/\Lambda^2)}{\log(-t+Q^2_0)/\Lambda^2}),
\label{tad}
\eeq
where
\beq
\alpha_s(k^2_t)=\frac{4\pi}{b\log(k^2_t/\Lambda^2)}, b=11-2N_f/3.
\label{b} \eeq
We included $Q_0^2$ in eq.~\ref{tad}  to match parton distributions
in the limit $t\rightarrow 0$. The same result can be obtained by
calculating  the integral in transverse
momentum plane in the whole  region of $k_t$ explicitly, combining
all three terms in the integrand together using the  Feynman
parametrization:
 \beq
 I=\int^1_0dk^2_t dx \frac{(\vec
k-(1-x)\vec \Delta)^2+t(1-x)^2}{(\vec k-(1-x)\vec \Delta)^2-tx(1-x)}.
\label{run1}
\eeq

\par
Finally let us note that we cannot distinguish in the LLA whether the
argument in the coupling constant is $k^2_t$ or $k^2$.  In the paper we
follow suggestion of ref. \cite{DMW}, that choice of $k^2_t$ minimizes
higher order corrections.

 \par
 The above results show that effectively the nonforward DGLAP distribution differs from
 the forward one only by the substitution of the variable
$\chi=\frac{1}{b}\log(\frac{\log(Q^2/\Lambda^2)}{\log(Q^2_0/\Lambda^2)})$
by $\chi'$ given by eq.~\ref{tad}.  Then the nonforward PDF
satisfy the evolution equation:
 \beq
{dG^B_A(x,Q_0^2,Q^2,t)\over d \log (Q^2-t)}={\alpha_s(Q^2-t)\over 4\pi}\sum
_C\int^1_0{dz\over z}\Phi^B_C(z)G^C_A({x\over z},Q_0^2,Q^2,t).
\label{7}
\eeq
Here $\Phi^B_C(z)$ are the standard DGLAP kernels, $G's$ are the
nonforward ladders, $A,B,C$ denote the parton species.  Eq.~\ref{7} is
valid for all massless partons $A,B,C$. The solution of this
equation can be obtained from the solution for the forward
distributions  by the substitution $\chi\rightarrow \chi'$.

\par
The expression obtained in the previous subsection for the single cell
in the approximation when the running of the coupling constant is
neglected coincides with the expression in ref. \cite{Bartels}. This is
because such DL terms are the same within DGLAP and BFKL
approaches. \cite{DGLAP2}. In difference from ref. \cite{Bartels}
we  include the running coupling constant within the DGLAP framework
and derive an evolution equation.

\subsection{Nonforward parton distributions  for the hard inelastic exclusive  diffractive
processes within the LO DGLAP approximation.}

\par In order to  evaluate the amplitude $f$  we need to solve eq.~\ref{7} for the case of scattering off a  parton. All  we need to do is to solve the
corresponding DGLAP evolution for zero $t$ and then substitute
$\chi\rightarrow \chi'$. The corresponding parton distributions
for forward case were derived  in refs.\cite{GW,DGLAP2,DDT}:  gluon
distributions in the gluon, $D^G_G(x,Q^2)$, and in quark $D^q_G(x,Q^2)$.
 Since $D^q_G\ll D^G_G$ at small $x$ \cite{DDT} we shall need only the $D^G_G$
function.
\par
As we already stressed above the pQCD evolution effects at small x are reduced
to the double logarithmic(DL) approximation. Indeed, it was shown in ref. \cite{Forte}
 that the PDFs  evaluated using the DL approximation for  gluon and quark
distributions gives very good description of data at HERA energies. Moreover the double
logarithmic approximation gives a good description of the parton distributions even at
relatively large $x_B \le 10^{-2}$  \cite{DDT}. As a result for the theoretical description of the small $x$ inelastic diffraction it should be
sufficient to use the double logarithmic expressions for parton distributions.
In order to stress that we work with a gluon distribution with a single parton boundary
conditions we shall denote this PDF as $D(x,Q^2,t)$ below.
\par
The simplest description of the parton distribution in the
DLA is achieved after the Mellin transform:
\beq
D(j,t,Q^2,Q^2_0)=\int^1_0 z^{j-1} D(x,t,Q^2,Q^2_0).
\label{Mel}
\eeq
The  solution of the relevant DGLAP equation has the form
\cite{DGLAP2,DDT}:
\beq
D(j,\chi)=\exp(\gamma(j)\chi),
\label{rel}\eeq
where
\beq
\gamma (j)=\frac{4N_c}{j-1}-a, \, a=(11/3)N_c+(2/3)N_f.
\label{dim}
\eeq
For our case $N_c=3,N_f=3$. The solution in the
$x,\chi$ space is obtained after the inverse Mellin transform:
 \beq
 xD(x,Q^2,t)=\frac{1}{2\pi i}\int_C dj
x^{-j}G(j,\chi).
\label{dl}
\eeq
The integration contour is chosen to the right of
the $j=1$ singularity and parallel to the imaginary
axis. For the final answer we need to carry the inverse Mellin
transform of eq.~\ref{rel} and to substitute $\chi\rightarrow \chi'$
as it was explained in the previous section:
\beq
xD(x,Q^2,Q^2_0,t)=8N_c\chi'I_1(u)/u, \label{dla}\eeq
 where \beq
u=\sqrt{16N_c\log(x/x_J)\chi'}.
\label{te}
\eeq
The function $I_1$ is the  modified Bessel function of the first kind. For $t=0$
this is just the formula for $D^G_G$ in the DLA, first obtained in ref.  \cite{GW}. For very small $x$ we
have the asymptotic expansion: \beq xD(x,\chi')=
\frac{\chi'^{1/4}N_c^{1/4}}{\sqrt{2\pi}\log(x_J/x)^{3/4}}\exp(u),\label{d5}\eeq
where  $x_J$ is given by eq.~\ref{int}.
\par
The formulae \ref{dla} and  \ref{d5} give the solution for nonforward parton distributions
that will be building blocks for the diffraction amplitude $f$.

\subsection{The nonforward parton distributions beyond the DGLAP approximation.}

\par
In the previous subsections we derived the formulae for the amplitude $f$  within the DGLAP approximation
(eq.\ref{dla}) which predicts the zero rate of the increase with energy for the nonforward PDF in
the kinematical range $-t\ge Q^2+M^2_V$. Formally we derived  this result within LO
DGLAP approximation.  But account of NLO, NNLO, ...approximations will not change this prediction
because all DL terms disappear in this kinematical domain.
\par
The DGLAP approximation ignores the possible contribution of
$\ln(x_0/x)$ terms that are  not enhanced by large
$\log(Q^2/Q^2_0-t)$ terms. These terms are connected with the gluon
radiation in the multi Regge kinematics. The BFKL and resummation
models \cite{BFKL,Salam,Ciafaloni} take into account both such
terms and the double logarithmic terms \cite{BFKL,DGLAP2}. However,
at present the direct numerical comparison between the BFKL and
DGLAP results is not possible since the BFKL based models do not
include the phase space constraints on the multi Regge gluon radiation,
that follow from the energy-momentum conservation, and in addition
neglect
the running of the coupling constant. Recently the attempt
was made to include such effects using the so called resummation
models approach \cite{Salam,Ciafaloni,ciafaloni1}. However, these were
models developed for $t=0$ only. The
interrelation
between the Pomeron behavior and the
double logarithmic approximation
in
the multi Regge kinematics needs an additional investigation.
 \par
 In order to take into account the terms that do not contain large logarithms
originating from  the integration over parton transverse momenta it is convenient to represent a parton distribution as
the product of DL contribution and the contribution that is not
included in DLA. We may parametrize the nonforward parton
distribution in the form \ref{f} as:
\beq f(x,Q^2,t)=f_{\rm
DL}(x,Q^2,t)F(x,t). \label{n1} \eeq \par
As we already mentioned, the Pomeron behavior due to emission of gluon in the multi Regge
 kinematics is delayed till very high energies as a consequence of  the strong constraints due to the local energy momentum conservation.

\par
Consider first the HERA energies.
  The interval in rapidity necessary  for
the emission of  the gluons adjacent in rapidity in the multi
Regge kinematics is at least $\sim 2\div 2.5$ \cite{FSXX,Schmidt,FRS}.
Since $\ln(x_0/x)\ge \ln(M^2_V/Q^2_0)$, the photon fragmentation
region occupies within the DLA  at least $\ge \log(M^2_V/Q_0^2)\gg 2\div 3$
units in rapidity. The proton dissociation in the triple reggeon
limit occupies at least three  units in rapidity (due to acceptance of the HERA detectors).
Rapidity span for  the kinematics of HERA is $\sim 8\div 9$ units. Thus in the HERA
kinematics it is hardly possible to emit even one additional gluon in the
multi Regge kinematics. This means that at HERA energies
if $-t\le Q^2+M^2_V$ there is no enough phase space for multi
Regge corrections, and the resulted energy dependence
is given the  DGLAP terms, i.e. $F=1$. For the
kinematic range $-t\ge Q^2+M^2_V$ the same kinematic analysis
shows that because of the diminishing of the photon fragmentation
region
due to disappearance of DL terms where maybe a room
for the radiation of one  gluon within
the multi Regge kinematics. We obtain:
\beq
F(x,t)=1+\beta(Y-\Delta Y)\theta (Y-\Delta Y).
\label{g11a}
\eeq
Here $\Delta Y\sim 4\div 5$ is the minimal rapidity needed for the
 start of a gluon radiation in a a multi Regge kinematics.  Numerically we
expect that the coefficient of the logarithmic term $\beta (t)\ll
\alpha_{\Pomeron}(t)-1$  since the existence  of large logarithmic  corrections
of this type at HERA energies would contradict to a good agreement between DL asymptotics and experiment in the
entire kinematic range of HERA \cite{Forte}. The intercept of Pomeron trajectory $\alpha_{\Pomeron}(t)$
is two - three  times smaller in the resummation  models,
than in the LO BFKL approximation because of the accurate account of the DL terms
\cite{Salam,Ciafaloni}.
Thus with a good accuracy we can put $F=1$   at HERA also for $-t\ge Q^2+M^2_V$.

\par
Consider now the energies significantly larger than the ones reached
at HERA.  In the ultraperipheral processes at the LHC one may
reach $x\sim 10^{-6}$, i.e. up to 10 units in rapidity available
for the HID process, after the exclusion of the proton fragmentation region, that
corresponds to 3 units in rapidity at least (see above). The good agreement between the
results of DGLAP and resummation models up to $x\sim 10^{-4}\div 10^{-5}$ for $Q^2\sim 30$ GeV$^2$
($\alpha_s\sim 0.2$) indicate that DL terms occupy the region of at least (2 $\div $3)$\log(Q^2/Q_0^2)$
in rapidity, i.e. $4 \div 7$ units for $Q^2\sim M^2_V\sim 10$ GeV$^2$ for charmonium production.
 The DL terms
define region in rapidity occupied by the photon fragmentation region and
result in a reduction of the rapidity interval corresponding to the   multi Regge
kinematics. This means that only $3\div 5$ units in rapidity at most may be
available for the emission of gluons in  the multi Regge kinematics $-$ less than the
 minimal region $\Delta Y\sim 5$. Thus in
the   ultraperipheral processes which could be studied at the LHC
for small $-t<M^2_V$,  our equations should be applicable, at least qualitatively.

For the kinematic range $-t\ge Q^2+M^2_V$ the double logarithms are absent.
In  this case the rapidity range available for the gluon emission could reach
$8\div 9$ units for multi Regge kinematics (after subtracting the photon fragmentation
 region (impact factor occupies $\sim 2$ units in rapidity).  Hence in this case  2 $\div $ 3 gluons could be emitted in the  multi Regge kinematics.

\par
Thus
the DL behavior will continue to reveal itself in the entire interval
of energies available at ultraperipheral collisions at the LHC. Only at the maximum energies available at the LHC
and $-t>M^2_V$ there is a sufficient phase space for the multi Regge gluon
emission.
In the kinematics of  HERA the energy dependence is
given by DGLAP approximation. For LHC energies, at the maximum energies to
be probed in ultraperipheral processes, and at $-t\sim M^2_V$ it will be
interesting to look for the Pomeron behavior which will compete with the onset of the black disk regime.

\section{The energy dependence of cross sections of  hard inelastic processes with proton dissociation.}

\par
Let us apply our results to the exclusive inelastic production of charmonium.  The QCD
factorization theorem  \cite{BFGMS,CFS}  allows us to
evaluate the diffractive cross section in terms of the convolution of nonforward PDF
discussed in the previous section with the parton distribution describing proton dissociation.
The corresponding Feynman graph are shown at  Fig.~1.  We shall argue that the obtained
above energy dependence of nonforward amplitude can
be used as the interpolation formula correct also for $-t\sim Q^2+M^2_V$.

\par
Let us consider first the kinematic range $Q_0^2\ll-t\ll
Q^2+M^2_V$. Since $-t\ll Q^2+M^2_V$  we may use the dipole
approximation \cite{BFGMS}. We expand the impact factor into
Taylor series over exchanged gluons transverse momenta  , and then
follow the steps used in the proof of QCD factorisation theorem.
Thus the amplitude $f$ of the process is proportional to \beq
f(-t,M_V,Q)=K \int^1_0 du\int d^2r (\phi_{V}
(u,r,\Delta)\triangle_t\phi_{\gamma}(r,u,\Delta)xD(x,4r^2,t).
\label{t1} \eeq The proportionality constant K is energy
independent. It is matrix in the space  of photon and vector meson
polarizations:  $L\rightarrow L$, $T\rightarrow T$,$L\rightarrow
T$, $T\rightarrow L$, where $L,T$ are longitudinally and
transversely polarized quarkonium and incident virtual photon.

$D$ is a nonforward PDF, discussed in the previous sections. The argument of
the nonforward parton distribution  $x$ is given by eq.~\ref{d1}:
\beq
x\sim (x_1+x_2)/2\sim 3(M^2_{V}+Q^2)/( 2x_Js).
\label{d1a}
\eeq
Let us stress that the only assumption used in the derivation of eq.~\ref{t1} is the possibility to approximate
impact factor by a first term in the Taylor expansion in exchanged gluon transverse momentum $q_1$.
 Numerically such approximation is rather effective.  The characteristic momenta of quarks
in the impact factor $r^2\sim (Q^2+M^2_V)/4$ (it is possible to  neglect
here the $t$-dependent term since $-t\ll Q^2+M^2_V$).
In addition $u\approx 1/2$ \cite{koepf}. This means that the expression \ref{d1a} can be
actually approximated as
\beq
f=xD((M^2_V+Q^2)/(2 x_Js),(M^2_V+Q^2),t)\Phi(t,Q^2,M^2_V),\label{t10}\eeq
where
\beq
\Phi (t,Q^2,M^2)\sim \int d^2r\int_0^1 du(\phi_{V}
(u,r,\Delta)\triangle_t\phi_{\gamma}(r,u,\Delta).
\label{t11}
\eeq
$\Phi$ depends on the  quarkonium wave functions and  influences the t-dependence of the
cross section but it is
independent of the incident energy.  The energy dependence of the diffraction
amplitude is entirely given by the convolution of nondiagonal forward distribution
$D((M^2_V+Q^2)/(2 x_Js),M^2+Q^2,t)$  described in detail in the previous section with a parton
distribution measured in DIS.  Thus in the kinematical range $Q_0^2\ll-t\ll Q^2+M^2_V$
the energy dependence of the cross section at fixed $x_j=-t/M^2_X$ is determined solely  by the function $D$ (eq.~ \ref{dla})
and does not depend on  details  of the photon and quarkonium wave functions.

\par
If momentum transfer $-t\sim Q^2+M^2_V$ is large, the
factorization theorem in the form described above should be
modified. This is because  the diffractive process is no more
dominated by the DGLAP kinematics  and the dipole approximation is
not valid.  The amplitude is still given by the convolution of the
impact factor and the gluon distribution. However in this
kinematic region there is only one transverse scale $-t$, and the
large logarithms originated due to QCD evolution are absent. Hence
in the DGLAP approximation all amplitudes are reduced to energy
independent impact factor.  One would obtain the same result if we would simply extrapolate
the energy dependence obtained above to this region.
Thus we can use the energy dependence given by eqs.
\ref{dla},\ref{n1} in the entire kinematic range $Q_0^2\le -t\le
Q^2+M^2_V$.  Beyond the DGLAP approximation  the gluon distribution
may contain the energy dependent terms which are
not enhanced by
large logarithms related to the $Q^2$ evolution. However it was argued
in the previous section that such terms can give only small
correction due to the energy-momentum constraints.

\par
We conclude that the energy dependence of the total diffractive
cross section in the DGLAP approximation can be explicitly calculable
in the kinematic range $-t\le Q^2+M^2_V$ as \beq
\frac{d\sigma}{dtdx_j}=\Phi(t,Q^2,M^2_V)^2\frac{(4N_c^2I_1(u))^2}{\pi
u^2}((81/16) G(x_J,t)+ \sum_i (q_i(x_J,t)+\bar q_i(x_J,t)).
\label{fe} \eeq Here
\begin{eqnarray}  u&=&\sqrt{16N_c\log(x/x_J)\chi'}, \, \chi^{'}=\frac{1}{b}\log(\frac{\log(Q^2/\Lambda^2)}{\log(-t+Q^2_0)/\Lambda^2}), \nonumber\\[10pt]
x_J&=&-t/(M^2_X-m^2_p-t),x\sim 3(Q^2+M^2_V)/(2x_Js), \, b=11-2/3N_f,N_c=3,\nonumber\\[10pt]
\label{fe1}
\end{eqnarray}
and $\Phi(t,Q^2,M^2_V)$ is the energy independent function,
which
depends on the
details of the wave functions of the  produced quarkonium and a photon.

\par
For the kinematic region $-t\sim  Q^2+M^2_V$ the cross section becomes energy independent
in the kinematics covered by HERA, up to logarithmic corrections of
the order $(1+ 2\beta(t)\log(x/x_J))$ for $\log(x_J/x)>4\div 5$, and $\beta (t)\ll
\alpha_{\Pomeron}(t) -1 $.

\par
The important case of charmonium photoproduction can be
obtained from the previous formulae  just by putting  $Q^2=0$.

To illustrate the pattern  of the variation with $t$ of the dependence of the cross section on the rapidity gap interval we present in Fig. 4
the logarithmic derivative $\displaystyle{\frac{d\log(d^2\sigma /dtdx_J)}{d\log(x/x_J)}}$  for the $J/\psi $ photoproduction at $-t= 0, 4, 9
$~GeV$^2$. It corresponds to the
effective "local"
value of $2\alpha_{\Pomeron}(t) - 2$. One can see that this quantity rapidly decreases with increase of $-t$. Note that the discussed
 approximation somewhat overestimates effective $Q^2$ for photoproduction
 leading to a somewhat stronger energy dependence than a more realistic analyses of the $t=0$ photoproduction of $J/\psi$ \cite{koepf}.

Hence we conclude that for a fixed $x_J$  and $-t \ge 4 $ GeV$^2$ the discussed process should depend very
weakly on the rapidity gap interval at the HERA energies. This is consistent with the recent
HERA findings \cite{galina} and phenomenological analysis of the earlier data  \cite{FSZ}.

It is worth mentioning that at sufficiently high energies where one
 may expect an increase of the amplitude due to the BFKL type dynamics
 the absorptive effects may strongly modify the energy dependence of the cross section. Really
 the process we consider is an example of the inelastic diffractive process. Such processes disappear
 in the black disk regime. In particular if $x_J$ is kept fixed and sufficiently large (say $\ge 0.05$) the
 contribution of the scattering at large impact parameters would be suppressed, while for
 the small impact parameters the probability of interaction with a gap in rapidity goes to zero.
  Since such effects strongly depend on the size of the $Q\bar Q $ dipole they would also result in a weaker dependence of the cross section on $t$.
These effects will be considered elsewhere.

\section{DVCS scattering.}

\par
The  knowledge  of the nonforward parton distribution evaluated above allows to calculate
dependence on energy of the process of diffractive photon production: $\gamma^*(Q^2)+p\to \gamma+p$
-DVCS.  Although this process is hardly possible to observe at large $t$   we present our results for
 completeness.

\par
Amplitude of this process is
\beq
\propto xG(x,Q^2,t)=F(t,x, Q^2)\, xG(x,\chi'),
\label{EBF}
\eeq
where $xG(x,\chi)$ is the gluon distribution  at HERA, whose energy
dependence is well approximated by \cite{Forte}
\beq
xG(x,\chi)\sim \exp(\sqrt{16N_c\log(x_0/x)\chi}), x_0\sim 0.1,Q_0^2\sim 1 {\rm GeV}^2,\label{gl}
\eeq
and $\chi'$ is given by eq.~\ref{tad}.
 \beq
 F_{2g}(t, x)=1/(1+t/m^2_g(x, Q^2))^2,
 \label{14}
 \eeq
is dipole approximation for the two gluon form factor of the nucleon
 \cite{FSZ1}. (We suppressed here the weak dependence of
 $F_{2g}$ on  $Q^2$).
  $F_{2g}(t,x)$ may depend on energy in a
restricted range of energies because of pion cloud around nucleon carrying small fraction of
proton momentum. So its contribution at $x\ge 0.1$ should be negligible,  But hard interaction
rapidly increases with  interaction  energy and at sufficiently small x pion tail should reveal itself
\cite{FSW}.

\par
The
expression
 \ref{EBF} had been derived in the impulse
approximation, which is
questionable at large $-t$.
However this effect is important for the calculation of absolute
value of amplitude but it will not change the energy dependence.

\section{Conclusions}

\par
We found different QCD dynamics  in the cross section of HID at HERA energies for fixed
$x_J=-t/M^2_X$  in the two
kinematical regions. In the first region $Q_0^2\le -t\le Q^2+M^2_V$ we calculated the energy
dependence of cross section within the DL  approximation - eq.~\ref{fe}.  In the second region
$-t\ge Q^2+M^2_V$ we showed that  the cross section of the HID processes is energy independent. This result
is  exact within the DGLAP approximation -valid  within any order of LO, NLO....NNLO  approximations.
Our calculations explains observed in the recent HERA data on the HID processes \cite{galina}  significant decrease
of the intercept of pQCD "Pomeron" with increase of $-t$ as compared to the intercept at t=0 .
The corrections  to DGLAP approximation due to gluon radiation in the multi-Regge kinematics
are small at the kinematics of HERA, as it follows both from the analysis of energy-momentum constraints on
the multi Regge gluon radiation, and
  the analysis of ref. \cite{Forte} that DLA gives a good description of the HERA data.
\par We argued that in ultraperipheral collisions
  at LHC the DL will give at least qualitatively good description
of energy dependence of HID processes in most of the kinematic
space available. However there may be a kinematic region $x\sim
10^{-6}, -t\ge M^2_V$, that lies in the borderline of available
energies where it will be very interesting to look for the onset
of the pomeron behavior.
\par
In this paper we focused  only on the energy dependence of the diffractive cross sections.
The detailed numerical studies of the cross sections, including their absolute values and t-dependence
will be presented elsewhere \cite{BFS}.

B.Blok is indebted to L.Lipatov for the illuminating discussion of
the properties of BFKL  and resummation models.

\newpage
\begin{figure}[htbp]
\centerline{\epsfig{figure=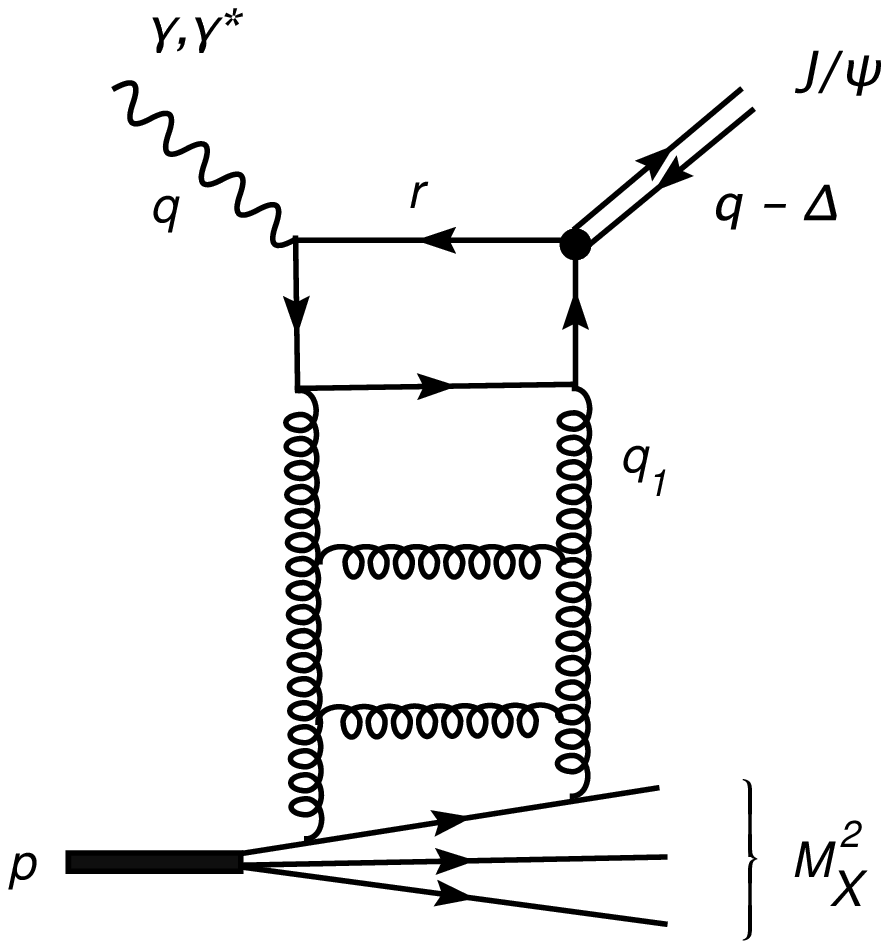,width=15cm,height=15cm,clip=}}
\caption{ The Feynman diagram describing the double diffractive
process in the tripple  reggeon limit in pQCD (there is also a cross
diagram, not depicted explicitly.)}\label{S1}
\end{figure}
\clearpage
\begin{figure}[htbp]
\centerline{\epsfig{figure=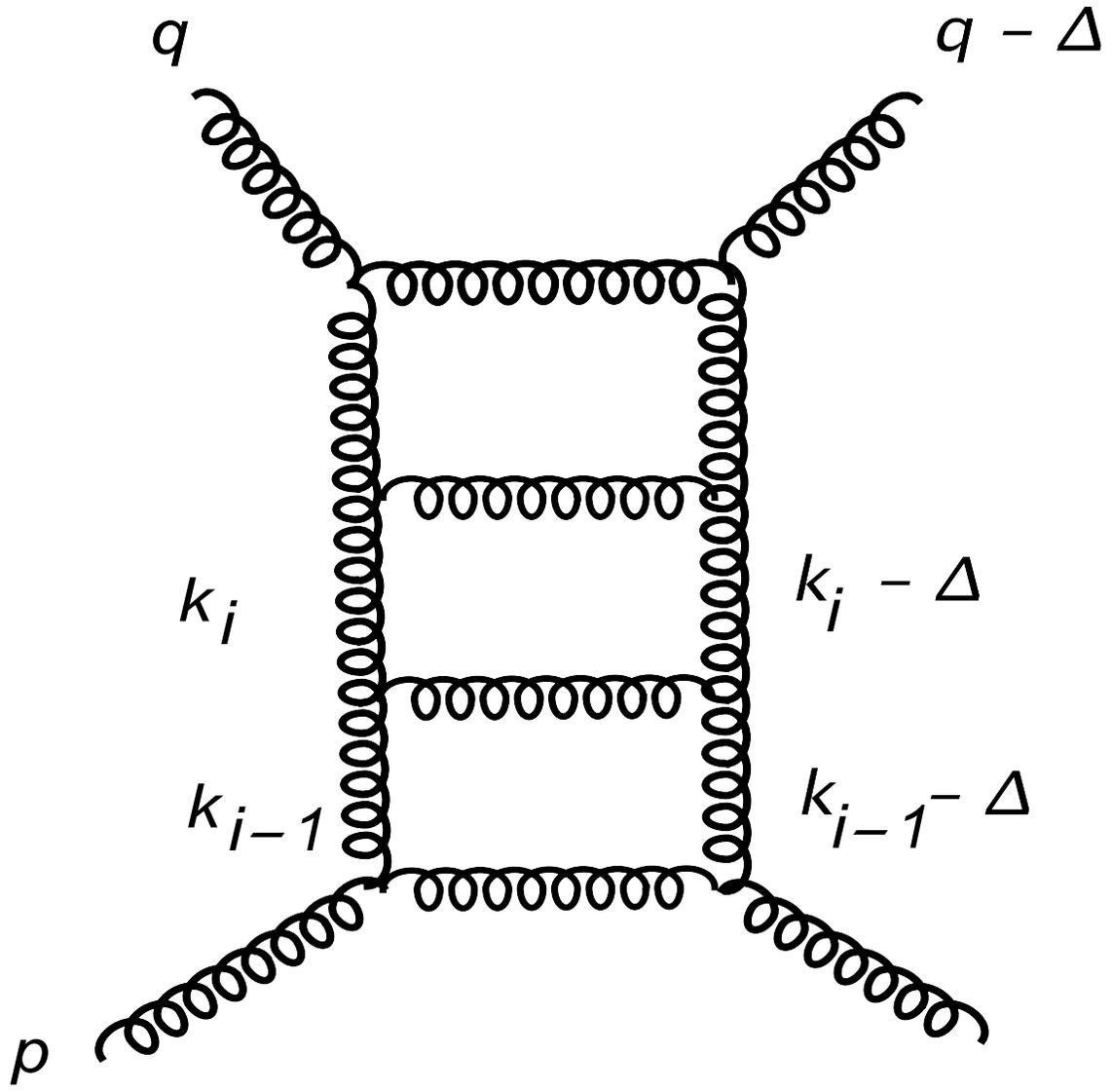,width=15cm,height=15cm,clip=}}
\caption{ The nondiagonal ladder}\label{S2}
\end{figure}
\clearpage
\begin{figure}[htbp]
\centerline{\epsfig{figure=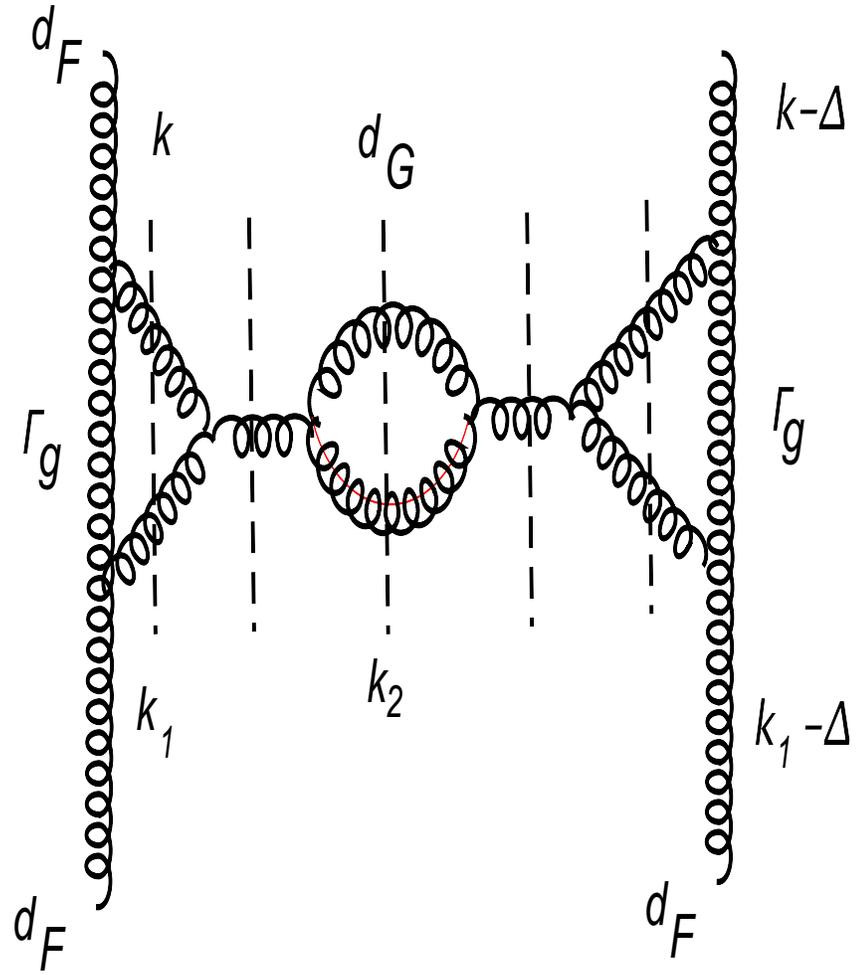,width=15cm,height=15cm,clip=}}
\caption{ The renormalization of the ladder.}\label{S3}
\end{figure}
\clearpage
\begin{figure}[htbp]
\centerline{\epsfig{figure=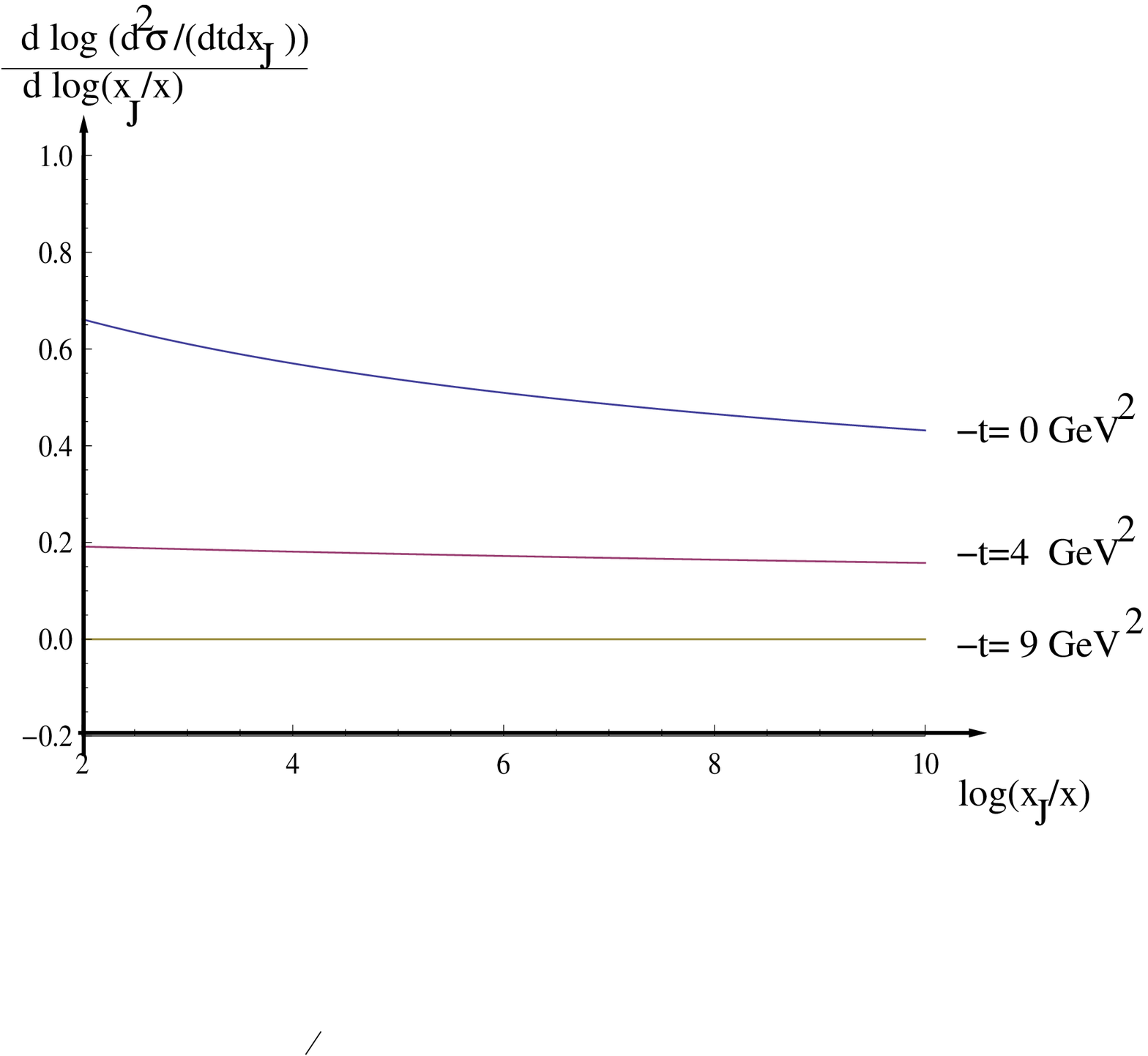,width=15cm,height=15cm,clip=}}
\caption{ The logarithmic derivative of the differential cross section as a function of energy for  $-t=0,4,9 $ GeV$^2$.}\label{S4}
\end{figure}

\clearpage

\newpage\begin{thebibliography}{}
\bibitem{ZEUS} M. Derrick et al. [ZEUS Collaboration] Phys. Lett. B 369 (1996)
55;S. Chekanov et al. [ZEUS Collaboration] Eur. Phys. J. C 26
(2003) 389.
 \bibitem{H1}[6] C. Adloff et al. [H1 Collaboration]
Eur. Phys. J. C 24 (2002) 517,  A. Aktas et al. [H1 Collaboration]
Phys. Lett. B 568 (2003) 205, A. Aktas et al. [H1 Collaboration]
Phys. Lett. B 638 (2006) 422.
\bibitem{galina} S. Chekanov et al, ZEUS collaboration, preprint
DESY-09-137, Measurement of $J/\psi$ photoproduction at large
momentum transfer at HERA.
\bibitem{AFS}H. Abramowicz, L. Frankfurt and M. Strikman, Surveys
in High Energy Physics, 11 (1997) 51,
also in proceedings SLAC Summer Institute, 1994, 539-574, SLAC, 1995.
\bibitem{Forshaw1}J. R. Forshaw, M.G. Ryskin,Z.Phys.C68 (1995) 137.
\bibitem{Levin}
E. Gotsman, E. Levin, U. Maor, E. Naftali, Phys.Lett.B532 (2002) 37 .
\bibitem{BFGMS} S. J. Brodsky, L. Frankfurt, J. F. Gunion, A. H. Mueller and
M. Strikman, Phys. Rev. D50 (1994) 3134.
\bibitem{CFS}John C. Collins,  Leonid Frankfurt, Mark Strikman,
 Phys.Rev.D56 (1997) 2982.
\bibitem{Mark} L. Frankfurt and M. Strikman, in Phys. Reports, 455 (2008) 105.
\bibitem{FSZ} L. Frankfurt, M. Strikman, M. Zhalov,Phys.Lett.B670 (2008) 32.
\bibitem{Lipatov} L. Lipatov,Sov.Phys.JETP 63 (1986) 904.
\bibitem{DGLAP1}V.N. Gribov and L. N. Lipatov, Sov. J. of Nucl. Phys.,
15 (1972) 438,672.
\bibitem{DGLAP2}Yu.L. Dokshitzer, Sov. Phys. JETP 46 (1977)
641.
\bibitem{DGLAP3} G. Altarelli and G.Parisi, Nucl. Phys., B126
(1977) 298.
\bibitem{Forte} R. Ball and  S.  Forte,Phys.Lett.B335 (1994) 77;
ibid, B351 (1995) 313;
ibid, B358 (1995) 365.
\bibitem{DDT} Yu. Dokshitzer,D. Diakonov and S. Troyan, Phys. Reports, 58  (1980)
269.
\bibitem{BFKL} E. Kuraev, V. Fadin, L. Lipatov, Sov. Phys.-JEP, 44
(1976) 443; 45 (1977) 199. I. Balitsky and L. Lipatov, Sov. J.
Nucl. Phys., 28 (1978) 822.
\bibitem{Ciafaloni}M. Ciafaloni, P. Colferai, G.P. Salam,
A. M. Stasto, Phys. Lett., 587 (2004) 87; Phys. Rev. D68 (2003)
114003.
\bibitem{ABF}G. Altarelli, S. Forte, R.D. Ball,
Nucl. Phys., B621 (2002) 359; B674 (2003) 459.
\bibitem{HERA} C. Adlov et al, Phys. Lett., B520 (2001) 183; S. Chekanov et al, Nucl. Phys. B713 (2005) 3.
\bibitem{transition}  M. Ciafaloni , D. Colferai , G.P. Salam , A.M. Stasto,
Phys.Rev.D66 (2002) 054014.
\bibitem{ciafaloni1} M. Ciafaloni, P. Colferai, G.P. Salam, JHEP 9910 (1999) 017.
\bibitem{lipatovfadin} L. Lipatov and V. Fadin,Phys.Lett.B429 (1998) 127.
\bibitem{FS2} L. Frankfurt,  A. Freund, V. Guzey, M. Strikman,
 Phys.Lett.B418 (1998) 345, Erratum-ibid.B429 (1998) 414.
\bibitem{Freund} A. Freund, V. Guzey, Phys.Lett.B462 (1999) 178.
\bibitem{Bartels} J. Bartels and A. Loewe,Z.Phys.C12 (1982) 263.
\bibitem{Gribov} V. Gribov, Theory of the complex momenta, Cambridge academic press 2003.
\bibitem{FS3} L. Frankfurt and M. Strikman, Phys. Rev. Lett, 63 (1989) 1914.
\bibitem{GM}V.N. Gribov, A. A. Migdal,  Yad.Fiz.8:1213,1968.
\bibitem{DMW} Y. Dokshitzer, G. Marchesini and G. R.
Webber,Nucl.Phys.B469 (1996) 93.
\bibitem{GW}D.J. Gross, F. Wilczek, Phys.Rev.D9 (1974) 980.
\bibitem{Salam} G. Salam,JHEP 9807 (1998) 019.
\bibitem{FSXX} L. Franfurt and M. Strikman, Nucl. Phys. Proc. Suppl. 79 (1999) 671.
\bibitem{Schmidt} C. Schmidt, Phys.Rev.D60 (1999) 074003.
\bibitem{FRS} J. R. Forshaw,  D. Ross, A. Sabio-Vera,  Phys. Lett. B455 (1999) 273.
\bibitem{koepf} W. Koepf, L. Frankfurt, M. Strikman, Phys.Rev.D54 (1996) 3194.
\bibitem{FSZ1}L. Frankfurt, M. Strikman and C. Weiss, Phys. Rev. D69 (2004)  114010.
\bibitem{FSW} L. Frankfurt, M. Strikman and C. Weiss, Ann.Rev.Nucl.Part.Sci.55 (2005) 403.
\bibitem{BFS} B. Blok, L. Frankfurt, M. Strikman, in preparation.

\end{thebibliography}
\end{document}